\documentclass{article}
\usepackage{color}
\usepackage[latin1]{inputenc}
\usepackage{amsmath}
\usepackage{amssymb}

\newcommand{\be}{\begin{equation}}
\newcommand{\ee}{\end{equation}}
\newcommand{\beann}{\begin{eqnarray*}}
\newcommand{\eeann}{\end{eqnarray*}}
\newcommand{\bea}{\begin{eqnarray}}
\newcommand{\eea}{\end{eqnarray}}

\newcommand{\bdm}{\begin{displaymath}}
\newcommand{\edm}{\end{displaymath}}

\usepackage[dvips]{graphicx}

\usepackage{epsfig}
\usepackage{graphicx}

\begin{document}
 
\centerline{\bf Conservation Laws in Gravitation and Cosmology}
\vspace{0.3cm}
\centerline{J\'{u}lio C. Fabris\footnote{fabris@pq.cnpq.br}}
\vspace{0.3cm}
\begin{center}
Departamento de F\'isica, Universidade Federal do Esp\'irito Santo, Avenida Ferrari 514,
Vit\'oria, 29075-910, Esp\'irito Santo, Brazil
\end{center}

\begin{abstract}
The existence of conservation laws is one of the most important requirement of physical theories. Some of them, like energy conservation, knows no experimental exception. However, the generalization of these conservation laws to curved space presents many challenges. The implementation of conservation laws in the General Relativity theory is revised, and the possibility of the generalization of the usual expression is discussed. The Rastall's theory of gravity, which considers a modification of the usual conservation of the energy-momentum tensor, is discussed in more detail. Some applications of the Rastall's theory to cosmology are presented, showing that it can lead to competetive results with respect to the Standard Cosmological Model.
\end{abstract}

\leftline{PACS: 04.50.Kd, 95.35.+d, 95.36.+x, 98.80.-k}

\section{Introduction}

The conservation laws are one of the cornestone of physics. Classical, Newtonian physics contains in its core the concepts of conservation of mass, momentum, angular momentum and energy. These concepts
have been enlarged in the electromagnetic theory, introducing in the conservation equations the quantities related to the fields, a framework that is not possible to include in context of the pure
Newtonian theory. The fact that the electromagnetic theory is, at the end, a relativistic theory, invariant by Lorentz transformations, led the association of the energy, momentum and angular momentum conservation laws as consequence of symmetries of the space-time. Such relation can be recast in much more general structure through the Noether's theorem which associates to each symmetry of a given theory
a conserved charge and, consequently, a conservation law: conservation laws express symmetries.
\par
Considerations of symmetries allow to determine the energy-momentum tensor in gauge and gravity theories. The energy-momentum tensor encodes all informations we have about
the energy, pressure and stress in a given physical system. In gauge theories, formulated in the Minkowski space-time, there is a canonical procedure
to obtain the energy-momentum tensor, in terms of the field content of the theory. In gravity the situation is, in principle, different since the geometry itself is dynamical containing its own degrees of freedom, and it is
not restrict to a passive r\^ole as the Minkowski space-time, the geometry of ordinary quantum field theory. In gravitational system, a metric energy-momentum tensor can be constructed, which is in principle equivalent to the canonical energy-momentum tensor concerning the fields of the system. A nice proof of this property is given, for a specific case, in reference \cite{fleming}. However, in very general cases, including torsion for example, such equivalence is not guaranteed. 
\par
The metric energy-momentum tensor (to employ the terminology of reference \cite{fleming}) is conserved when a Riemannian geometry is considered. This conservation law is directed connected with the
invariance of the theory with respect to coordinate transformation. However, such conservation of the energy-momentum tensor must be seen with caution. In a gravitational background, there will be
inevitably an exchange of energy between the field components and the gravitational field, represented by the metric. Hence, the conservation laws in geometric theories of gravitation have not exactly
the same sense as in Minkowski space-time. This fact has led to many generalizations of the usual conservation laws of the metric energy-momentum tensor.
\par
We revise, in this text, the status of the conservation law in geommetric theories of gravity. Some cosmological examples are considered. The generalization of usual conservation law are reviewed, with
special emphasis to the that proposed by the Rastall theory of gravity. Interesting applications of this theory to
the dark matter/dark energy problem in cosmology today are shown.   

\section{Definitions of the energy-momentum tensor}

In the Newtonian physics, the conservation laws come directly from the Newton's equations and from the imposition that matter is not created. All these informations are encode in the continuity equation and
in the Euler's equation, that is, the Newton second law adapted to the hydrodynamic context, supplemented by the Poisson equation which describes the gravitational potential:
\begin{eqnarray}
\frac{\partial\rho}{\partial t} + \nabla\cdot(\rho\vec v) &=& 0,\\
\frac{\partial \vec v}{\partial t} + \vec v\cdot\nabla \vec v &=& - \frac{\nabla p}{\rho} - \nabla\phi,\\
\nabla^2\phi &=& 4\pi G\rho,
\end{eqnarray}
where $\rho$ is the density, $p$ is the pressure, $\vec v$ the velocity field and $\phi$ the gravitational potential. Matter, momentum, energy conservations law are implicit in these equations. Dissipative
process may be taken into account through the generalization of the Euler's equation, leading to the Navier-Stokes equation.

The extension to relativistic dynamics is somehow straighforward, taking the form in four dimensions given by,
\begin{eqnarray}
\partial_\rho J^\rho &=& 0,\\
m_0\frac{d^2 x^\mu}{d\tau^2} &=& F^\mu,
\end{eqnarray}
where $J^\mu$ is the four-current, $F^\mu$ is the four-force, $\tau$ is the proper time and $m_0$ is the rest mass. However, all the momentum, energy, stress content is encoded in the energy-momentum tensor,
$T_{\mu\nu}$ from which those quantities can be obtained.

Let us concentrate first in the usual definitions of the energy-momentum tensor. We can follow the reasoning of reference \cite{landau}. For simplicity, we will consider scalar fields only, which can
represent, in a more general situation, the degrees of freedom of the system. Initially, we have in mind fields defined in Minkowski space-time. The action is given generally by
\begin{equation}
{\cal A} = \int {\cal L}(\phi,\partial\phi)d^4x.
\end{equation}
Variation with respect to $\phi$ and $\partial_\rho\phi$, leads to the usual Euler-Lagrangian equations,
\begin{equation}
\partial_\rho\frac{\partial{\cal L}}{\partial\phi_{,\rho}} - \frac{{\cal L}}{\partial\phi} = 0.
\end{equation}
It is shown in the reference \cite{landau} that the quantity
\begin{equation}
T_{\mu\nu} = {\cal L}\delta_{\mu\nu} - \phi_{,\mu}\frac{\partial{\cal L}}{\partial^{,\nu}\phi},
\end{equation}
is conserved:
\begin{equation}
\partial_\mu T^{\mu\nu} = 0.
\end{equation}
In brief, this is the canonical prescription. However, the energy-momentum defined in this way is not unique. In fact, it is possible to add to it a quantity $\partial_{\lambda}G^{\lambda\mu\nu}$,
where $G^{\lambda\mu\nu}$ is anti-symmetric in $\mu$ and $\lambda$, so that $\tilde T^{\mu\nu} = T^{\mu\nu} +  \partial_{\lambda}G^{\lambda\mu\nu}$ is also conserved. Moreover, such arbitrariness has
as consequence that energy-momentum tensor is not necessarily symmetric. In reference \cite{landau}, the symmetry and unicity is obtained through the definition of the momentum, from which a conserved angular
momentum is constructed.  A similar reasoning is employed in reference \cite{ryder}, where it is explicitly established that angular momentum conservation leads to a unique and symmetric energy-momentum tensor.
\par
The energy momentum-tensor can be also constructed using the metric as a dynamical degree of freedom. Going back to the action in an arbitrary geometric background, we have
\begin{equation}
{\cal A} = \int {\cal L}(\phi,\partial\phi)\sqrt{-g}d^4x.
\end{equation} 
Imposing a coordinate transformation $x^\mu \rightarrow x^\mu - \xi^\mu$, we have,
\begin{equation}
g_{\mu\nu} \rightarrow g_{\mu\nu} + \xi_{\mu;\nu} + \xi_{\nu;\mu}.
\end{equation}
Defining \cite{landau}\footnote{This definition is, in fact, a direct application of the Euler-Lagrange equations.},
\begin{equation}
\label{def1}
T_{\mu\nu} = \frac{2}{\sqrt{-g}}\biggr\{\partial_\rho\frac{\partial(\sqrt{-g}{\cal L})}{\partial g^{\mu\nu}_{,\rho}} - \frac{\partial(\sqrt{-g}{\cal L})}{\partial g^{\mu\nu}}\biggl\},
\end{equation}
we find, after some manipulations and discarding a surface term,
\begin{equation}
\delta{\cal A} = \int {T^{\mu\nu}_{;\mu}}\xi_\nu\sqrt{-g}d^4x.
\end{equation}
Since, $\xi_\nu$ is arbitrary it implies, considering a stationary condition for the action,
\begin{equation}
\label{c-l}
T^{\mu\nu}_{;\mu} = 0,
\end{equation}        
with $T^{\mu\nu}$ defined by (\ref{def1}). Defined in this way, the energy-momentum tensor is automatically symmetric and conserved.
\par
In reference \cite{fleming}, the equivalence between the metric and canonical derivations of the energy-momentum tensor is established in the absence of spinors fields, up eventually to the adition
of an anti-symmetric term, as discussed previously. The important point to remark is that the invariance with respect to coordinate transformation is directly connected with the conservation of the
energy-momentum tensor in its metric formulation.

\section{Cosmological implications}

In this section we will resctrict ourselves to the usual conservation law, given by (\ref{c-l}), which plays a central r\^ole in constructing the General Relativity theory. In fact,
the Einstein's equations, with a cosmological constant $\Lambda$,
\begin{equation}
\label{e-e}
R_{\mu\nu} - \frac{1}{2}g_{\mu\nu}R - g_{\mu\nu}\Lambda = 8\pi GT_{\mu\nu},
\end{equation}
have this form precisely in order to have (\ref{c-l}) through the Bianchi's indentity. Of course, Einstein's equations equation have other distinguinshing features that make them very special.
They lead to a set of non-linear second order differential equations, what implies that the Cauchy problem is naturally well-posed. Second, it can be derived from a Lagrangian that it is the most
general Lagrangian leading to second order differential equations: the Einstein-Hilbert Lagrangian is, in fact, the two first terms of the Lovelock complete series of the most general geometrical Lagrangian 
leading to second order differential equations \cite{love}, the other terms of the series not contributing to the field equations in four dimensions.
\par
All these features seem to single out the relation (\ref{c-l}) as the correct expression for the conservation laws in a geometric gravity theory. Even if all this seems very convincing, some curious
consequences emerge from (\ref{c-l}) revealing particular features concerning the usual conservation law in Newtonian and relativistic (non-gravitational) conservation law. One example comes direct from
the cosmological applications of (\ref{c-l}).
\par
Let us consider the Friedmann-Lema\^{\i}tre-Robertson-Walker space-time, given by the isotropic and homogeneous metric,
\begin{equation}
\label{flrw}
ds^2 = dt^2 - a(t)^2\biggr\{\frac{dr^2}{1 - kr^2} + r^2(d\theta^2 + \sin^2\theta d\phi^2)\biggl\},
\end{equation}
where $a(t)$ is the scale factor describing the dynamics of the universe, and $k$ is the (constant) curvature of the spatial section.
For the energy-momentum tensor, we write is as a direct generalization to curved space-time of the usual energy-momentum tensor for a fluid in special relativity:
\begin{equation}
T_{\mu\nu} = (\rho + p)u_\mu u_\nu - pg_{\mu\nu},
\end{equation}
where $\rho$ is the energy density, $p$ is the pressure and $u_\mu$ is the vour velocity of the fluid.
The complete set of equations emerging from the Einstein's equations (\ref{e-e}) and from the conservation law (\ref{c-l}), with a zero cosmological constant, are
\begin{eqnarray}                                                                                                                                                                                                \biggr(\frac{\dot a}{a}\biggl)^2 + \frac{k}{a^2} &=& \frac{8\pi G}{3}\rho,\\
2\frac{\ddot a}{a} + \biggr(\frac{\dot a}{a}\biggl)^2 &=& - 8\pi G p,\\
\dot\rho + 3\frac{\dot a}{a}(\rho + p) &=& 0.
\end{eqnarray} 
These three equations are not independent due to the Bianchi's identity. We have just to function to determine, $a$ and $\rho$, since $p$ indicates which kind of fluid it is being considered,
what mounts out to fix an equation of state, an external ingredient with respect to this set of equations.
\par
Let us consider a simple, particular case. It is given by imposing a spatially flat universe (which is, by the way, suggested by observations \cite{komatsu}) and a linear barotropic equation
of state $p = \omega\rho$, with $\omega =$ constant. Then, the equations reduce to:
\begin{eqnarray}
\biggr(\frac{\dot a}{a}\biggl)^2 &=& \frac{8\pi G}{3}\rho,\\
\dot\rho + 3\frac{\dot a}{a}(1 + \omega)\rho &=& 0.
\end{eqnarray}
These equations have the well-known solutions:
\begin{equation}
a(t) \propto t^\frac{2}{3(1 + \omega)}, \quad \rho \propto a^{-3(1 + \omega)}.
\end{equation}
If we have worked in a Newtonian framework, we will have obtained $a(t) \propto t^{2/3}$ and $\rho \propto a^{-3}$. The interpretation of the Newtonian results is quite simple: the behaviour of
the density just express the diluting of the gas of particles of mass $m$ due to the expansion of the universe. The Newtonian solutions correspond to the case $\omega = 0$ in General Relativity case.
This is expected: since the universe is homogeneous and isotropic, there is no pressure gradient and, consequently, no pressure effect.
\par
But, in General Relativity framework pressure plays a r\^ole even with the high symmetry of the space-time. If $\omega > 0$, density decreases faster than in the Newtonian situation as the universe
expand; on the other hand, if $\omega < 0$\footnote{Negative pressure is an allowed concept even in usual thermodynamics, see reference \cite{landaubis}.}, density decreases slower than in the Newtonian case. If $\omega = - 1$, the density becomes constant with the expansion of the universe.
Moreover, if the null energy condition limit is violated ($\omega < - 1$), density grows with the expansion of the universe, a very counterintuitive situation. 
\par
We can interpret this strange behaviour (from the Newtonian point of view) in two ways:
\begin{enumerate}
\item In General Relativity pressure {\it weights}, leading to an increase in the attractive behaviour if $\omega > 0$, and leading to repulsive contribution if $\omega < 0$.
This is somehow a reminmiscence of the famous special relativity relation $E = mc^2$: all forms of energy contribute to the gravitational mass.
\item There is an exchange of energy between matter and the gravitational field. This is represented by the work done by pressure. In some sense, this mounts out to a way back to the 
ancient
notion of gravitational field, which has been replaced by the notion of geometry. Such proposal may be re-obtained by a modification of the Newton's conservation law, implemented by the
so-called {\it neo-newtonian theory} \cite{neo}.
\end{enumerate}
\par
In any case, there is a departure from the usual Newtonian framework, and conservation law in this cosmological context must been seen in more large context.
\par 
There is another aspect of the conservation law (\ref{c-l}) that must be remarked. The null divergence concerns the total energy-momentum tensor. When there are multiple fluids the separate conservation
of each component is a choice, that may be dictated by physical considerations. For example, in presence of radiation and baryonic matter, it is mandatory that each of these components must obey
(\ref{c-l}) separately. But, there is evidence of a dark sector in the universe, with dark matter (a pressureless component) and dark energy (a component displaying negative effective pressure).
In this case, since these components are unknown, there is no a priori reason to ignore the possibility that there is a direct exchange of energy between
the two components. Hence considering that the total energy-momentum tensor is a sum of two components, $T^{\mu\nu}_T = T^{\mu\nu}_1 + T^{\mu\nu}_2$.
\begin{equation}
{T^{\mu\nu}_T}_{;\mu} = 0 \Rightarrow {T^{\mu\nu}_1}_{;\mu} = - {T^{\mu\nu}_2}_{;\mu}.
\end{equation}
Such interacting model is a very active line of research about the dark sector of the universe. One of the reasons is that, as already stated above, the observations indicate the existence of
a dark sector in the Universe, with two components, dark matter and dark energy. Each of these components have scales very differently with the expansion of the universe. However, they have today a
very similar value. This is called the {\it the cosmic coincidence problem}. While this is, apparently, a {\it coincidence} in the Standard Cosmological Model, it can have a dynamical explanation if
the two components exchange energy between them, leading even to possibility that the present acceleration of the universe is a transient phenomena, driven essentially by the interaction in the dark sector \cite{winfried,nelson}.

\section{Some non-conservative theories of gravity}

In 1949, in a short paper 
by Jordan in the {\it Nature} \cite{jordan}, it was considered the possibility that the creation of matter could not imply a violation of the conservation of energy. A rough evaluation reveals that the potential energy
of the universe is of the same order than the energy related to the total rest mass. Hence, matter creation could occur but keeping the total energy the same, since the positive energy associated to
the rest mass of the created particle is compensated by negative potential energy associated with the interaction of the created particle with all remaining mass of the universe.
\par
One year before, the idea of cosmological models satisfying the perfect cosmological principle has been presented \cite{bondi,hoyle}. According to the perfect cosmological principle, not only the universe has the same
appearence in all directions (isotropy) and positions (homogeneity), but also in all times. This generalization of the usual cosmological principle implies that the density must be constant in time.
And, if the universe is expanding, the creation of matter is unavoidable.
\par
As a matter of fact, the perfect cosmological principle is compatible with an expanding universe only if the Hubble function is a constant for all times - the expansion rate is time independent.
This means $\frac{\dot a}{a} = H = H_0 =$ constant. Hence, $a \propto e^{H_0 t}$, i.e., we must live in a de Sitter universe. A de Sitter universe implies, in principle, an equation of state of the type
$p = - \rho$, since, for this case,
\begin{equation}
\dot\rho + 3\frac{\dot a}{a}(\rho + p) = 0 \quad \rightarrow \quad \dot\rho = 0 \quad \rightarrow \quad \rho = \mbox{constant}.
\end{equation}
This would imply that the universe would have just the vacuum component, what is not a realistic scenario. However, keeping the same equations we can give another interpretation for conservation law.
The conservation equation can be re-written as,
\begin{equation}
\dot\rho + 3\frac{\dot a}{a}\rho = \Gamma, \quad \Gamma = 3\frac{\dot a}{a}\rho.
\end{equation}
Here, $\Gamma$ is re-interpret as the matter creation rate. Hence, the universe is matter dominated but with matter creation, leading to an effective pressure which is equivalent to
vacuum equation of state. This is an example of a situation where ${T^{\mu\nu}}_{;\mu} \neq 0$, if only the pressureless component is considered. The reinterpretation as a conserved vacuum energy
is possible, but may be just a formal device to recover the usual expression for the conservation of the energy-momentum tnesor. This is just an example that frequently
a non-conserved fluid with a given equation of state may imply conserved fluid with an effective equation of state and, in some sense, the derivation sketched above for the energy-momentum tensor may be somehow kept untouched, but
intepreting it as a derivation of this effective energy-momentum tensor. We will come back to similar situations later.
\par
Such idea of a stationary universe has been revived later, in the Hoyle-Narlikar theory \cite{h-n}, where a new field have been introduced in order to take into account properly of the matter creation
mechanism. The idea is to consider an action formulation for a gravity and a minimally coupled scalar field, called $C$ \footnote{For this reason, such theory has also been called {\it $C$-field theory}}:
\begin{equation}
{\cal A} = \int\biggl\{\frac{R}{16\pi G} - \frac{1}{2}C_{;\rho}C^{;\rho}\biggl\}\sqrt{-g}d^4x - \sum_a m_a\int ds + \sum_a\int C_{;\rho}ds^\rho.
\end{equation}
The first two terms of this action would mean just the Einstein-Hilbert theory with a scalar field as the matter source. The third term is just the geodesic action. The fourth term is the crucial
mechanism for the matter creation: the world-line of a particle suffers the action of the scalar (creation) field, leading to the possibility of production of matter. Taking into account the
matter component and the creation field, the Einstein's equation are now,
\begin{eqnarray}
R_{\mu\nu} - \frac{1}{2}g_{\mu\nu}R &=& 8\pi G(T^m_{\mu\nu} + T^C_{\mu\nu}),\\
{T^{\mu\nu}_m}_{;\mu} &=& - {T^{\mu\nu}_C}_{;\mu}.
\end{eqnarray}
Moreover,
\begin{equation}
\Box C = \bar n,
\end{equation}
where $\bar n$ is the number of particles created by unit of proper volume. From the equations above, we can see that this matter creation theory is an example where the overall General Relativity scenario
is preserved but predicting that only the {\it total} energy-momentum tensor is conserved.
Remark that in order to have matter creation, it is necessary that the energy of the $C$-field is negative: the creation of positive mass is made at the expense of the increasing of the negative energy
of the $C$-field. In principle, such property may lead to instability at quantum level. But, it has been arqued that the counter-reaction on the field equations would avoid the system to fall in
an infinite large negative state.
A detailled discussion of these matter creation theories can be found in reference \cite{narlikar}.  
\par
The stationary cosmological scenario predicted by the matter creation theories are highly disfavoured when compared with observations. Even if they explain the expansion of the universe,
the abundances of light chemical elements (predicted in a natural way in the big-bang scenario) asks for some deep intelectual exercices to be acquainted in the context of the matter
creation theories. The same occurs for the Cosmic Background Radiation, $CMB$. Hence, the interest for such theories has decreased exponentially the last decades. However, recently the
idea of a non-trivial coupling between a scalar field and matter has been revived in a different context, and with different purposes, some of them already discussed above.
\par
Let us consider now a non-miminal gravity and scalar field coupling. The action reads,
\begin{equation}
\label{bd}
{\cal A} = \int\biggl\{\phi R - \tilde\omega\frac{\phi_{;\rho}\phi^{;\rho}}{\phi}\biggl\}\sqrt{-g}d^4x + \int {\cal L}_m\sqrt{-g}d^4x.
\end{equation}
In this action, we have change the scalar field notation from $C$ to $\phi$ in order to forget (at least for the moment) the matter creation idea. Morevover, we have written $\frac{1}{16\pi G} = \phi$,
and we have introduced a free dimensionless constant in the kinetic term of the scalar field, $\tilde\omega$. Action (\ref{bd}) represents in fact the Brans-Dicke theory \cite{bd}. In this theory, matter is conserved. The final
field equations are the following:
\begin{eqnarray}
R_{\mu\nu} - \frac{1}{2}g_{\mu\nu}R &=& \frac{8\pi}{\phi}T_{\mu\nu} + \frac{\tilde\omega}{\phi^2}\biggl(\phi_{;\mu}\phi_{;\nu} - \frac{1}{2}g_{\mu\nu}\phi_{;\rho}\phi^{;\rho}\biggr)\nonumber\\
&+& \frac{1}{\phi}\biggl(\phi_{;\mu;\nu } - g_{\mu\nu}\Box\phi\biggl),\\
\Box\phi &=& \frac{8\pi T}{3 + 2\tilde\omega},\\
{T^{\mu\nu}}_{;\mu} &=& 0.
\end{eqnarray} 
In the Brans-Dicke theory, the energy-momentum tensor is conserved. But, the results of the theory depend strongly on the value of the dimensionless parameter $\tilde\omega$. The local tests implies a very high 
value of $\tilde\omega$, reducing the theory essentially to the General Relativity framework, with a constant $\phi$ \cite{will}. On the other hand, the most interesting theoretical perspective offered by the Brans-Dicke
theory is given for $\tilde\omega$ small or even negative \cite{brasil}.
\par
A way to surmount the difficulties imposed by the local test is to rewrite the Brans-Dicke theory through a conformal transformation.
In fact, imposing a transformation of the type
\begin{equation}
g_{\mu\nu} = \phi^{-1}\tilde g_{\mu\nu},
\end{equation}
the conservation law takes now the form,
\begin{equation}
\tilde \nabla_\mu \tilde T^\mu_\nu = - \frac{1}{2}\frac{\phi_{;\nu}}{\phi}\tilde T.
\end{equation}
Now, the energy-momentum tensor is not conserved anymore, and there is a direct exchange of energy between the scalar field and matter. 
This may lead to the so-called {\it chameleon mechanism}, since now the constraints on the behaviour of the scalar field (hence, the experimental estimations for $\omega$) may depend on the
matter environement. 
This may allow to reconcile the local tests, requiring huge values of $\tilde\omega$, with the large scale tests, for which a small (or even negative) value of $\tilde\omega$ is favoured. Such mechanism may also be implemented in the context of $f(R)$ theories \cite{felice}.                        

\section{Rastall's theory}

In the beginning of the seventies, Rastall proposed a modification of the gravity theory \cite{rastall}, where the energy-momentum tensor $T^{\mu\nu}$ does not obey the usual conservation law. Instead, the divergence of $T^{\mu\nu}$ reads
\begin{equation}
{T^{\mu\nu}}_{;\mu} = \kappa R^{;\nu}\;,
\end{equation}
where $R$ is the Ricci scalar and $\kappa$ is a constant. The field equations in Rastall's theory read
\begin{eqnarray}
R_{\mu\nu} - \frac{1}{2}g_{\mu\nu}R &=& 8\pi G\left(T_{\mu\nu} - \frac{\gamma - 1}{2}g_{\mu\nu}T\right)\;,\\
{T^{\mu\nu}}_{;\mu} &=& \frac{\gamma - 1}{2}T^{;\nu}\;,
\end{eqnarray}
where $\gamma$ is a dimensionless constant connected to $\kappa$. When $\gamma = 1$, General Relativity theory is recovered.
\par
The modification of the energy-momentum tensor law proposed by Rastall is based on the remark that the usual conservation law has been tested only in Minkowskian limit. Moreover, the curved background
leads to interpretation problems, as discussed above. But Rastall's theory has a major conceptual difficult to surmount: in principle, it has not a Lagrangian formulation. In fact, this is a controversial
aspect of the theory. First, to which extend a Lagrangian formulation is required in order a given theory to be considered physical? It is clear that the Lagrangian formulation allows the identification
of the symmetries. But, some symmetries can exist even in non-Lagrangian theory. For example, a theory of the kind,
\begin{equation}
R_{\mu\nu} - \frac{1}{2}g_{\mu\nu}R + \beta R_{\mu\rho}R^\rho_\nu = 8\pi G T_{\mu\nu},
\end{equation}
obeys the general covariance even if it does not come from a Lagrangian.
\par
Even though, an action formulation of the Rastall's theory seems to be possible, by using a non-canonical measure, what mounts to implement the theory in a non-Riemannian geometry framework \cite{smalley}.
The Weyl's geommetry opens some perspective in this sense. 
\par
Rastall's theory could be seen as a redefinition of the energy-momentum tensor. In fact, if we define
\begin{equation}
\tilde T_{\mu\nu} = T_{\mu\nu} - \frac{\gamma - 1}{2}g_{\mu\nu}T,
\end{equation}
we recover General Relativity, with $\tilde T_{\mu\nu}$ as the matter source. In this sense, Rastall's theory could be seen as a procedure to generate effective, sometimes {\it exotic}, equations of state from the usual ones \footnote{Such interpretation of Rastall's theory is due to the late Patricio Letelier.}. 
\par
In fact, considering the cosmological background and an equation of state of the type $p = \omega\rho$, we obtain the effective equation of state \cite{kerner},
\begin{equation}
\omega_{eff} = \frac{(5 - 3\gamma)\omega + \gamma - 1}{3 - \gamma - 3(1 - \gamma)\omega}.
\end{equation}
Remark that the vacuum energy equation of state $\omega = - 1$, implies a $\omega_{eff} = - 1$: it is a kind of fixed point.
\par
However, this interpretation is valid for a single fluid model with constant equation of state. In more general situations, the scenario can be more complex, as it will be seen below. Moreover,
the modification of the conservation law proposed by the Rastall's theory lead to many interesting new perspective to the analysis of gravitational systems.
One example, is the self-interacting scalar field formulation, as it will be discussed now.
\par
Let us consider the canonical form for the energy-momentum tensor of a self-interacting scalar field, i.e.
\begin{equation}
T_{\mu\nu} = \phi_{,\mu}\phi_{,\nu} - \frac{1}{2}g_{\mu\nu}\phi_{,\rho}\phi^{,\rho} + g_{\mu\nu}V(\phi)\;,
\end{equation}
Inserting it in the Rastall's equation obtain the following coupled equations:
\begin{eqnarray}
\label{Ein00} R_{\mu\nu} - \frac{1}{2}g_{\mu\nu}R &=& \phi_{,\mu}\phi_{,\nu} - \frac{2 - \gamma}{2}g_{\mu\nu}\phi_{,\alpha}\phi^{,\alpha} + g_{\mu\nu}(3 - 2\gamma)V(\phi)\;,\\
\Box\phi + (3 - 2\gamma)V_{,\phi} &=& (1 - \gamma)\frac{\phi^{,\rho}\phi^{,\sigma}\phi_{;\rho;\sigma}}{\phi_{,\alpha}\phi^{,\alpha}}\;.
\end{eqnarray}
From Eq.~\eqref{Ein00}, the following effective energy-momentum tensor can be read off:
\begin{equation}
\label{efetivo}
T_{\mu\nu}^{eff} = \phi_{,\mu}\phi_{,\nu} - \frac{2 - \gamma}{2}g_{\mu\nu}\phi_{,\alpha}\phi^{,\alpha} + g_{\mu\nu}(3 - 2\gamma)V(\phi)\;,
\end{equation}
implying the following expressions for the energy density and pressure in cosmological background:
\begin{equation}
\rho_\phi^{eff} = \frac{\gamma}{2}\dot\phi^2 \quad, \quad p_\phi^{eff} = \frac{2 - \gamma}{2}\dot\phi^2 - (3 - 2\gamma)V(\phi)\;.
\end{equation}
Using this expression in to evaluate the speed of sound, given by,
\begin{equation}
c_s^2 = \frac{p_\chi}{\rho_\chi}, 
\end{equation}
where $\chi = \frac{\dot\phi^2}{2}$ is the usual kinetic term,
one finds
\begin{equation}
c_s^2 = \frac{\gamma - 2}{\gamma}\;.
\end{equation}
This implies a vanishing speed of sound for $\gamma = 2$. In this case, the non-canonical self-interacting scalar field based on Rastall's theory may represent dark matter \cite{liddle}. Such possibility
does not exist in the canonical context, since for a canonical self-interacting scalar field we have $c_s^2 = 1$. On the other side, from the non-perturbative point of view, the "Rastall's scalar field" can represents dark energy by a suitable choice of the potential. We will explore this possibility in the section 7.
\par
Before to develop an application of the non-canonial scalar field predicted by the Rastall's theory, we must call attention to one property of this formulation. If we consider a single fluid
model, the perturbation of this non-canonical scalar field is consistent only if $\gamma = 1$ or if the fluctuations are homogeneous. In the first case, the theory reduces to General Relativity;
in the second case, the fluctuations are just a redefinition of the background. A consistent perturbation of the non-canonical scalar field is possible, on the other hand, if besides the non-canonical
scalar field matter is present. This is ok, since matter exists! A more detail discussion on this question can be found in reference \cite{oliver}.

\section{Rastall's comology and the $\Lambda$CDM model}

In order to investigate the possible observational status of the Rastall's theory, let us first consider how it can fit the Stantdard Cosmological Model.
\par
According to observations, the universe today is dominated by two exotic components, dark matter and dark energy. Dark matter is responsable for about $25\%$ matter-energy of the universe, and
dark energy accounts for $70\%$. Hence, the amount of known types of matter and energy (radiation, neutrino, baryons) is only $5\%$ of the total matter-energy content. Dark matter must have
an effective zero pressure, in order to play a r\^ole in the formation of structure, while dark energy must have negative pressure, in order to drive the accelerated expansion of the universe
today, and not agglomerate (significantly, at least) locally. The main candidates to describe dark matter are axions and neutralinos \cite{bertone} which are, however, hypothetical particles until now. The most
natural candidate for dark energy is the vacuum energy which theoretically has a value to large compared with that indicated by observations \cite{padma,caldwell}. All these results imply the acceptance of General 
Relativity as the correct theory of gravity. The model implement in the context of General Relativity employing the concepts of dark matter and dark energy is called the $\Lambda$CDM model. It is very succesfull at least at the background and linear perturbative level.
\par
How it is possible to implement a similar model in context of non-conservative theories of gravity? To do this let us consider the Rastall's theory. The $\Lambda$CDM is essentially a model for recent
periods of the cosmic evolution, where radiation and neutrino plays a subdominant r\^ole. Hence, let us consider a two fluid model, one with zero pressure ($p_m = 0$) and the other obeying the
vacuum energy equation of state ($p_x = - \rho_x$).
The total energy-momentum tensor is given by $T^{\mu\nu} = T^{mu\nu}_m + T^{\mu\nu}_x$. It must obey the Rastall's relation,
\begin{equation}
{T^{\mu\nu}}_{;\mu} = \frac{\gamma -1}{2}T^\nu.
\end{equation}
How to split this equation into two equations, one for each component?
One possibility is to impose that $T^{mu\nu}_m$ conserves in the usual way. The reason for that is the necessity for these components to agglomerate in order to form structures, as it happens
in the $\Lambda$CDM.
Hence, the final equations, following this ansatz, are,
\begin{eqnarray}
R_{\mu\nu} - \frac{1}{2}g_{\mu\nu}R &=& 8\pi G\biggr\{T^m_{\mu\nu} + T^x_{\mu\nu} - \frac{\gamma - 1}{2}g_{\mu\nu}(T^m + T^x)\biggl\},\\
\label{c1}
{T^{\mu\nu}_x}_{;\mu} &=& \frac{\gamma - 1}{2}(T_m + T_x)^{;\nu},\\
{T^{\mu\nu}_m}_{;\mu} &=& 0.
\end{eqnarray}
Remark that in the "conservation law" for the "dark energy" component the trace of the total energy-momentum tensor appears in the right hand side. If $\gamma = 1$, the equations of the
$\Lambda$CDM model are recovered.
\par
Let us impose a flat Friedmann-Lema\^{\i}tre-Robertson-Walker metric (\ref{flrw}). The equations of motion becomes the following:
\begin{eqnarray}
\label{f}
H^2 &=& \frac{8\pi G}{3}\biggr\{(3 - 2\gamma)\rho_x + \frac{-\gamma + 3}{2}\rho_m\biggl\},\\
\dot\rho_m + 3H\rho_m &=& 0,\\
(3 - 2\gamma)\dot\rho_x &=& \frac{\gamma - 1}{2}\dot\rho_m.
\end{eqnarray}
We have the following solutions for the mass densities:
\begin{eqnarray}
\rho_m &=& \frac{\rho_{m0}}{a^3},\\
\label{fbr2}
\rho_x &=& \frac{\rho_{x0}}{3 - 2\gamma} + \frac{\gamma - 1}{2(3 - 2\gamma)}\rho_m,
\end{eqnarray}
where we have written the integration in a particular way for future convenience.
Inserting in the "Friedmann's equation" (\ref{f}), we obtain,
\begin{equation}
H^2 = \frac{8\pi G}{3}(\rho_{x0} + \rho_m),
\end{equation}
which is the same equation we find in the usual $\Lambda$CDM model. The spatial componente of the field equations leads to,
\begin{eqnarray}
2\frac{\ddot a}{a} + \biggr(\frac{\dot a}{a}\biggl)^2 = 8\pi G \rho_{x0},
\end{eqnarray}
As a consequence, the background metric is {\it exactly} the same as in the $\Lambda$CDM model.
\par
However, there is a striking difference. Now, the dark energy component is given by equation (\ref{fbr2}), with a time-dependent behaviour (due to the non-homogenous term in
(\ref{c1})), while in the $\Lambda$CDM model the dark energy component is strictly constant - there is just the first term in (\ref{fbr2}) as it can be verified by imposing $\gamma = 1$.
\par
The fact that this Rastall's cosmological model reproduces the dynamics of the $\Lambda$CDM model at background level assures that all kinematics observational tests (those employing just
the non-perturbed metric) fit very well the Rastall's model also. But, what about the perturbative level? Let us consider only linear perturbations, and employ the synchronous coordinate condition.
After a long calculation, we find a unique equation for the the density contrast for the matter component, $\delta_m = \frac{\delta\rho_m}{\rho_m}$:
\begin{equation}
\ddot\delta_m + 2\frac{\dot a}{a}\delta_m - 4\pi G\rho_m\delta_m = 0.
\end{equation}
This is the same equation as the that found for the matter density contrast in the $\Lambda$CDM model! Hence, all linear perturbative observational test based on linear perturbations - for which the
$\Lambda$CDM model have good results - are well fitted.
But, again, there is a difference: now there is perturbations in the cosmological term, given by,
\begin{equation}
\delta_x = \frac{\gamma - 1}{2}\frac{\delta_m}{\frac{\rho_{x0}}{\rho_{m0}}a^3 + (\gamma - 1)}.
\end{equation}
Hence, dark energy now agglomerates, while in the $\Lambda$CDM model it remains strictly homogeneous.
This fact may have important consequences at non-linear level, a regime where the $\Lambda$CDM faces many difficulties.
\par
For more details on this model, see reference \cite{davi}.

\section{Rastall's scalar field and the Chaplygin gas model}

Such features of the Rastall's non-canonnical scalar field can be used to implement a scalar formulation of the Chaplygin gas  \cite{jackiw,moschella,berto1,neven}, as it was proposed for example
in reference \cite{thais}, curing some observational tensions existing
in the usual formulation of this unified model for dark matter and dark energy \cite{colistete,berto2,finelli,piattella1,hermano}.
\par
The generalized Chaplygin gas (GCG) model proposes to unify dark matter and dark energy into a single fluid obeying the equation of state,
\begin{equation}
p = - \frac{A}{\rho^\alpha},
\end{equation}
where $A$ and $\alpha$ are (constants) free parameters. The usual conservation law, leads
to
\begin{equation}
\rho(a) = \biggr\{A + B\,a^{-3(1 + \alpha)}\biggl\}^\frac{1}{1 + \alpha}.
\end{equation}
In general, the GCG model interpolates a matter dominated phase and a de Sitter phase. However, while the background observational tests (e.g., supernova type Ia) favors negative values of $\alpha$
\cite{colistete},
the perturbative analysis is restricted in the usual formulation to $\alpha > 0$ in order to keep the squared speed of sound positive \cite{hermano}.
\par
A canonical scalar reformulation of the GCG model can not cure this observational tension, since a canonical scalar field has a speed of sound given by $c_s^2 = 1$, and can not play the r\^ole
of dark matter in the past. But, this scalar representation seems possible in the context of Rastall's theory.
\par
In fact, considering a self-interacting scalar field, with the effective energy-momentum tensor (\ref{efetivo}), with $\gamma = 2$, we obtain,
\begin{equation}
T_{\mu\nu}^{eff} = \phi_{,\mu}\phi_{,\nu} -  g_{\mu\nu}V(\phi)\;.
\end{equation}
Inserting the FLRW metric, we have the following density and pressure associated with this scalar field:
\begin{equation}
\rho_\phi = \dot\phi^2 - V(\phi)\;, \quad p_\phi = - V(\phi)\;.
\end{equation}
Let us suppose that this density and pressure reproduce the background behaviour of the GCG model.
\par 
Using the effective expressions written above for the Rastall's scalar field (with $\gamma = 2$), we have,
\begin{eqnarray}
\dot\phi(a) &=& \sqrt{3\Omega_{c0}}\sqrt{g(a)^{1/(1 + \alpha)} - \bar A g(a)^{-\alpha/(1 + \alpha)}}\;,\\
V(a) &=& 3\Omega_{c0}\bar A g(a)^{-\alpha/(1 + \alpha)}\;,
\end{eqnarray}
where $g(a) \equiv \bar A + (1 - \bar A)a^{-3(1 + \alpha)}$. Hence, in order to have a zero speed of sound, the scalar model must obey the following equations:
\begin{eqnarray}
 R_{\mu\nu} &-& \frac{1}{2}g_{\mu\nu}R = 8\pi GT_{\mu\nu} + \phi_{,\mu}\phi_{,\nu} + g_{\mu\nu}V(\phi)\;,\\
 \Box\phi &+& V_\phi + \frac{\phi^{,\rho}\phi^{,\sigma}\phi_{;\rho;\sigma}}{\phi_{,\alpha}\phi^{,\alpha}} = 0\;,
\end{eqnarray}
where we have made the redefinition $V(\phi) \rightarrow - V(\phi)$. 
\par
Let us inspect now the perturbative behaviour of this system, computing scalar perturbations in the density contrast.
The perturbed equations in the synchronous coordinate condition read \cite{thais}:
\begin{eqnarray}
\label{rastall1}
\ddot\delta &+& 2\frac{\dot a}{a}\dot\delta - \frac{3}{2}\frac{\Omega_0}{a^3}\delta =
\dot\phi\dot\Psi - V_\phi\Psi\;,\\
\label{rastall2}
2\ddot\Psi &+& 3\frac{\dot a}{a}\dot\Psi + \left(\frac{k^2}{a^2} + V_{\phi\phi}\right)\Psi = \dot\phi\dot\delta\;,
\end{eqnarray}
where $\Psi = \delta\phi$ and $\delta$ is the density contrast of the matter component. Using the scale factor as independent variable, the above system of equations take the following form:
\begin{eqnarray}
 \delta'' &+& \left[\frac{2}{a} + \frac{f'(a)}{f(a)}\right] \dot\delta - \frac{3}{2}\frac{\Omega_0}{a^3f^2(a)}\delta =
\phi'\Psi' - \frac{V_\phi}{f^2(a)}\Psi\;,\\
2\Psi'' &+& \left[\frac{3}{a} + 2\frac{f'(a)}{f(a)}\right]\Psi' + \left[\frac{k^2}{a^2f^2(a)} + \frac{V_{\phi\phi}}{f^2(a)}\right]\Psi = \phi'\delta'\;,
\end{eqnarray}
where $f(a) = \dot a = \sqrt{\Omega_{m0}a^{-1} + \Omega_c(a)a^2}$ and $\Omega_c(a) = \Omega_{c0}g(a)^{1/(1 + \alpha)}$.
\par
\begin{figure}[htbp]
\begin{center}
\includegraphics[width=0.5\linewidth]{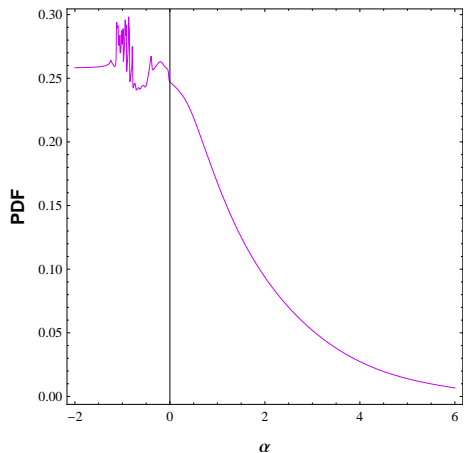}
\caption{One dimensional PDF for the parameter $\alpha$ in the Rastall's scalar model in presence of the baryonic component.}
\label{rastall-fig2}
\end{center}
\end{figure}

Using a Bayesian analysis and comparing the theoretical predictions of our model with the 2dFRGS data for the power spectrum of matter distribution in the universe, we find a significant probability region 
for $\alpha < 0$. The results are shown in figure \ref{rastall-fig2}, considering the unified scenario where, besides the GCG, there is only the baryonic component. To obtain the one-dimensional
PDF, the other degrees of freedom were marginalized (integrated). For more details, see reference \cite{thais}.
In conclusion, if the GCG model is represented by a non-canonical scalar field, like the one suggested by Rastall's theory of gravity, 
the observational tension that plagues the GCG fluid model may disappear or, at least, be considerably alleviated. This fact may open new perspectives for the dark matter-dark energy unification program.

\section{Conclusions}

Conservation laws play a fundamental r\^ole in physics. But, while in a
Newtonian framework the meaning of these conservatoin laws is clearly stated, some difficulties appear at the level of the
geometric theories. In curved space-time even the notion of gravitational energy is not universally accpeted.
In this sense, the implementation of conserved laws, mainly those related to the energy-momentum tensor admits many generalizations with respect to the usual expressions, inspired in the
flat space-time context.
\par
We have revised in this text the usual notion of the energy-momentum tensor conservation law. Later we have presented some possible generalizations. The first one is related to the notion
of matter creation in an expanding universe. 
Later, the formulation of scalar-tensor theory (e.g., Brans-Dicke) in terms of an interacting model for matter and the scalar field was mentioned. Such kind of interacting model has received
a lot of attention recently due to the possibility of reformulate the $f(R)$ theories in terms of the scalar-tensor theories. The interaction, in this case, allows to implement the chameleon
mechanism, giving the possibility to obtain very interesting results for very large scales and reproduce, at the same time, the local tests conveniently.
\par
Special attention has been given to the Rastall's theory of gravity. This theory, proposed in the beginning of the seventies, was introduced in view of the ambiguity of the notion of conservation
law in curved space-time.
The Rastall's theory gives very interesting results, at least at cosmological level, and some of its configurations may keep the success of the $\Lambda$CDM model, but adding new ingredients that
may have impacts at non-linear level, the regime where the $\Lambda$CDM model has many difficulties. Of course, Rastall's theory must follow a long way in order to become a serious alternative to
General Relativity. But, the results reported here may motivate a deep analysis of this theoretical framework.
\par
The many challenges represented by the generalization of the notion of energy and conservation laws to curved space-time reveal the necessity to keep a open-mind attitude concerning new
ideas in physics. What may seem "heretic" in a first moment may reveal (or not!) to be a fruitfull concept in the development of physics as a science. This has always been stimulated by
prof. Mario Novello, combining such open-mind attitude with scientific rigour, necessary to have solid scientific advances.

\section*{Acknowledgements}

Most of the results exhibited here have been obtained in works made in collaboration with A.B. Batista, C.E.M. Batista, M.H. Daouda, B. Fraga, O.F. Piattella, N. Pinto-Neto, R. de S\'a Ribeiro, D.C. Rodrigues and
W. Zimdahl. I thank A.B. Batista, O.F. Piattella, D.C. Rodrigues and W. Zimdahl for their carefull reading of the text, their criticisms and suggestions.
We acknowledge the partial financial support by CNPq (Brazil).



\begin{thebibliography}{99}
\bibitem{fleming} H. Fleming, Rev. Bras. Ens. Fis; {\bf 16}, 48(1994).
\bibitem{landau} L. Landau and E. Lifchitz, {\bf Th\'eorie du champ}, \'Editions Mir, Moscou(1967).
\bibitem{ryder} L.H. Ryder, {\bf Quantum field theory}, Cambridge university press, Cambridge(1996).
\bibitem{love} D. Lovelock, J. Math. Phys. {\bf 12}, 498(1971); {\bf 13}, 874 (1972).
\bibitem{komatsu} E. Komatsu et al., {\it Seven-year Wilkinson microwave anisotropy probe (WMAP) observations: cosmological interpretation}, arXiV:1001.4538.
\bibitem{landaubis} L. Landau and E. Lifchitz, {\it Physique statistique}, \'Editions Mir, Moscou(1967).
\bibitem{neo} W.H. McCrea, Proc. R. Soc. London {\bf 206}, 562 (1951);
E.R. Harrison, Ann. Phys. (N.Y.) {\bf 35}, 437 (1965).
\bibitem{winfried} S. del Campo, J.C. Fabris, R. Herrera and W. Zimdahl, Phys. Rev. {\bf D83}, 123006(2011).
\bibitem{nelson} J.C. Fabris, B. Fraga, N. Pinto-Neto and W. Zimdahl, JCAP, {\bf 011}, 008(2010.
\bibitem{jordan} P. Jordan, Nature {\bf 164}, 637(1949).
\bibitem{bondi} H. Bondi and T. Gold, Month. Not. R. Astron.
Soc. {\bf 108}, 252(1948).
\bibitem{hoyle} F. Hoyle, Month. Not. R. Astron. Soc. {\bf 108}, 372(1948).
\bibitem{h-n} F. Hoyle and J.V. Narlikar, Proc. Ro. Soc. {\bf A270}, 334(1962).
\bibitem{narlikar} J.V. Narlikar, {\bf Introduction to cosmology}, Cambridge university press, Cambridge(1983).
\bibitem{bd} C. Brans and R.H. Dicke, Phys. Rev. {\bf 124}, 925(1961).
\bibitem{will} C.M. Will, {\bf Theory and experiment in gravitational physics}, Cambridge university press, Cambridge(1993).
\bibitem{brasil} A.B. Batista, J.C. Fabris and R. de Sá Ribeiro, Gen. Rel. Grav. {\bf 33}, 1237(2001).
Subjects: General Relativity and Quantum Cosmolog
\bibitem{felice} A. de Felice and S. Tsujikawa, {\it $f(R)$ theories}, arXiv:1002.4928.
\bibitem{rastall} P. Rastall, Phys. Rev. {\bf D6}, 3357 (1972).
\bibitem{smalley} L. L. Smalley, Il Nuovo Cim. {\bf B80}, 42(1984).
\bibitem{kerner} J.C. Fabris, R. Kerner and J. Tossa, Int. J. Mod. Phys. {\bf D9}, 111(2000).
 \bibitem{liddle}  C. Gao, M. Kunz, A.R. Liddle and D. Parkinson, Phys. Rev. {\bf D81}, 043520(2010).
 \bibitem{oliver} J.~C.~Fabris, M.~H.~Daouda and O.~F.~Piattella,  Phys. Lett. {\bf B711}, 232(2012).
 \bibitem{bertone}  G. Bertone, D. Hooper and J. Silk, Phys. Rep. {\bf 405}, 279(2005). 
 \bibitem{padma}  T. Padmanabhan, Phys. Rep. {\bf 380}, 235(2003). 
\bibitem{caldwell} R.R. Caldwell and M. Kamionkowski, Ann. Rev. Nucl. Part. Sci. {\bf 59}, 397(2009).
\bibitem{davi} C.~E.~M.~Batista, M.~H.~Daouda, J.~C.~Fabris, O.~F.~Piattella and D.~C.~Rodrigues,  Phys. Rev.{\bf  D85}, 084008(2012).
 \bibitem{jackiw} R. Jackiw, {\it A particle field theorist's lectures on supersymmetric, non abelian fluid mechanics and d-branes},
physics/0010042. 
 \bibitem{moschella} A.Y. Kamenshchik, U. Moschella and V. Pasquier, Phys. Lett.
{\bf B511}, 265(2001); J.C. Fabris, S.V.B. Gon\c{c}alves and P.E. de Souza, Gen. Rel. Grav. {\bf 34}, 53(2002). 
\bibitem{berto1}
M.C. Bento, O. Bertolami and A.A. Sen, Phys. Rev. {\bf D66},
043507 (2002).
\bibitem{neven}  N. Bilic, G.B. Tupper and R.D. Viollier, Phys. Lett. {\bf B535}, 17(2002).
\bibitem{thais} J.C. Fabris, M. Hamani Daouda, T.C. Guio and O.F. Piattella, Grav\&Cosm. {\bf 17}, 259(2011).
\bibitem{colistete} R. Colistete Jr, J. C. Fabris, S.V.B. Gon\c{c}alves and P.E. de Souza, Int. J. Mod. Phys. D13,
669(2004); R. Colistete Jr., J. C. Fabris and S.V.B. Gon\c{c}alves, Int. J. Mod. Phys. D14,
775(2005); R. Colistete Jr. and J. C. Fabris, Class. Quant. Grav. 22, 2813(2005).
\bibitem{berto2}  T. Barreiro, O. Bertolami and P. Torres, Phys. Rev. {\bf D78}, 043530(2008).
\bibitem{finelli}  L. Amendola, F. Finelli, C. Burigana and D. Carturan, JCAP {\bf 0307}, 005(2003).
\bibitem{piattella1} O. Piattella, JCAP {\bf 1003}, 012(2010). 
\bibitem{hermano} J.C. Fabris, S.V.B. Gon\c{c}alves, H.E.S. Velten and W. Zimdahl, Phys. Rev. D78, 103523
(2008). J.C. Fabris, H.E.S. Velten and W. Zimdahl, Phys. Rev. D81, 087303(2010).


\end{thebibliography}
\end{document}